\documentclass[12pt,letterpaper]{article}
\pdfoutput=1
\usepackage{jheppub}
\usepackage{amssymb}
\usepackage{color}
\usepackage{slashed}
\usepackage{amsmath}
\usepackage{amsfonts}
\usepackage{amssymb}
\usepackage{graphicx,subcaption}
\usepackage{caption}
\usepackage{subcaption}
\usepackage{float}
\usepackage{tikz}
\usetikzlibrary[arrows]
\usepackage{amsmath}
\usepackage{epsfig}
\usepackage{appendix}
\usepackage{listings}
\usepackage{xcolor}
\usepackage{mathabx}

\newcommand{\be}{\begin{equation}}
\newcommand{\ee}{\end{equation}}

\newcommand{\tr}{{\rm tr\,}}

\title{On the planar free energy of matrix models}

\author{Bartomeu Fiol}
\author{and Alan Rios Fukelman}

\affiliation{Departament de F{\'\i}sica Qu\`antica i Astrof\'isica i \\Institut de Ci{\`e}ncies del Cosmos, 
Universitat de Barcelona,
Mart{\'\i}\ i Franqu{\`e}s 1, 08028 Barcelona, Catalonia, Spain}

\emailAdd{bfiol@ub.edu}
\emailAdd{ariosfukelman@icc.ub.edu}

\abstract{In this work we obtain the planar free energy for the Hermitian one-matrix model with various choices of the potential.
We accomplish this by applying an approach that bypasses the usual diagonalization of the matrices and the introduction of the eigenvalue density, to directly zero in the evaluation of the planar free energy. In the first part of the paper, we focus on potentials with finitely many terms. For various choices of potentials, we manage to find closed expressions for the planar free energy, and in some cases determine or bound their radius of convergence as a series in the 't Hooft coupling.  In the second part of the paper we consider specific examples of potentials with infinitely many terms, that arise in the study of ${\cal N}=2$ super Yang-Mills theories on $S^4$, via supersymmetric localization. In particular, we manage to write the planar free energy of two non-conformal examples: SU(N) with $N_f<2N$, and ${\cal N}=2^*$.}

\begin{document}
\maketitle
\section{Introduction}


In this work we study the planar limit of Hermitian matrix models with a single Hermitian N$\times$N matrix $\phi$, 
\be
\bar Z(g_s,c_i, c_{ij}) = \frac{Z(g_s,c_i,c_{ij})}{Z(g_s,0,0)}= \frac{ \int d\phi \, e^{-\frac{1}{2g_s} \tr \phi^2+ V(\phi)}}{\int d\phi \, e^{-\frac{1}{2g_s} \tr \phi^2}} \, ,
\ee
with a potential given by an arbitrary number of single- and double-trace terms of even powers of $\phi$,
\be
V(\phi)= \text{N} \sum_{i}c_{2i} \tr \phi^{2i}+\sum_{i,j}c_{2i\, 2j} \tr \phi^{2i} \tr \phi^{2j} \, ,
\label{def:MM_potential}
\ee
where the sums can contain either a finite or an infinite number of terms and the coefficients $c_{2i}, c_{2i\, 2j}$ are N-independent and arbitrary. The particular case where the potential has just a finite number of single-trace terms and no double-trace terms has been extensively studied for its relevance to various problems, from graph enumeration \cite{Bessis:1980ss} to two-dimensional quantum gravity \cite{Ginsparg:1993is, DiFrancesco:1993cyw}. Models with a finite number of double-trace terms in the potential have also been considered in the  context of two-dimensional quantum gravity \cite{Das:1989fq}. On the other hand, potentials of the form (\ref{def:MM_potential}) with an infinite number of terms appear in statistical physics, in Chern-Simons theories with matter, or in duals to M-theory on certain backgrounds as reviewed in \cite{Grassi:2014vwa}. More recently, it has been realized \cite{Billo:2019fbi, Fiol:2020bhf, Fiol:2020ojn, Fiol:2021icm} that supersymmetric localization \cite{Pestun:2007rz} reduces the evaluation of certain observables of four dimensional ${\cal N}=2$ super Yang-Mills (SYM) theories on $S^4$ to matrix models that can be recast to take the form of  (\ref{def:MM_potential}).

One of the basic objects of interest in the study of these matrix models is the free energy
\be
{\cal F} =F-F_{\text{gaussian}}= \log \bar Z= \log Z-\log Z_{\text{gaussian}}
\ee
and since the groundbreaking work \cite{Brezin:1977sv} it has been understood that one can organize the perturbative evaluations of the free energy in a double series expansion
\be
{\cal F}=\sum_{g\geq 0} \textnormal{N}^{2-2g} {\cal F}_g(t) =\sum_{g\geq 0} \textnormal{N}^{2-2g} \sum_{h\geq 0} F_{g,h} t^h 
\label{freefdef}
\ee 
where $t=g_s N$ is the 't Hooft coupling of the matrix model. Over the years, many powerful methods have been developed to evaluate these expansions for various potentials (see \cite{Marino:2004eq, DiFrancesco:2004qj, Eynard:2015aea, Anninos:2020ccj} for reviews). It is however fair to note that there are preciously few potentials for which closed forms for the free energy are known. A first question is then whether one can obtain a closed form expression for the free energy - or its terms in the 1/N expansion - for different choices of the potential. A second question that (\ref{freefdef}) immediately raises is the convergence of the perturbative series that appear. In Quantum Field Theory, the exponential growth of the number of Feynman diagrams at every order in the perturbative expansion implies that the default expectation is that, barring unusual cancellations, perturbative series are not convergent, but asymptotic. On the other hand, in the 1/N expansion (\ref{freefdef}), the number of diagrams at a fixed order in 1/N is drastically reduced, so it grows only as a power law \cite{Koplik:1977pf}. This opens up the possibility that these series have a finite radius of convergence \cite{Koplik:1977pf, tHooft:1982uvh, Garoufalidis:2010ye}.

The purpose of this work is to present some progress on  these two questions for various choices of potentials in (\ref{def:MM_potential}). In the first part of this work, we focus on potentials with a finite number of terms, and for various examples we find a closed expression for the planar free energy ${\cal F}_0(t)$ as an all-order perturbative series in the 't Hooft coupling. In some of these cases, we either determine or bound the radius of convergence of the resulting series. In the second part of the paper we consider very particular instances of potentials with infinitely many single- and double-trace terms. The potentials we consider are relevant for the study of four dimensional ${\cal N}=2$ SYM theories via supersymmetric localization \cite{Pestun:2007rz}. In particular, we provide the first examples of evaluation of ${\cal F}_0(t)$ for four dimensional non-conformal  ${\cal N}=2$ SYM theories. In this second part we don't present any new result on the convergence of the resulting series; nevertheless, we suggest that the results in the first part of the paper may play an important role in deriving analytically the radius of convergence found numerically in the literature.

To arrive at the results announced above, we have followed an approach that bypasses the more usual route of diagonalizing the matrix $\phi$ and introducing a density of eigenvalues, to directly zero in the planar free energy. This approach relies on two key ingredients. The first one is the explicit form of the connected planar $n-$point function \cite{tutte, Gopakumar:2012ny},
\be
\langle \text{tr } \phi^{2k_1}  \dots \text{tr } \phi^{2k_n}  \rangle_c = \frac{(d-1)!}{(d-n+2)!} \prod_{i=1}^n \frac{ (2k_i)!}{(k_i-1)! k_i!} t^d \textnormal{N}^{2-n} \, ,
\label{mm:cc}
\ee
where $d=\sum_i k_i$. In the particular case when all the $k$s are the same, (\ref{mm:cc}) reduces to
\be
\langle \left( \text{tr }\phi^{2k} \right)^n \rangle_c =  \frac{(nk-1)!}{(n(k-1)+2)!} \left(\frac{(2k)!}{k! (k-1)!} t^k\right)^{n} \, .
\label{samepowers}
\ee
The second ingredient, which builds upon the first, is a recently found combinatorial expression for the planar free energy of matrix models with potentials of the form (\ref{def:MM_potential}) \cite{Fiol:2020bhf, Fiol:2021icm} as a sum over a particular type of graphs, known as tree graphs
\be
\begin{split}
{\cal F}=\sum_{m=1}^\infty \frac{(-1)^m}{m!} \sum_{k=0}^m {m \choose k} &\sum_{p_1,\dots,p_{m-k}} c_{p_1}\dots c_{p_{m-k}} \sum_{\substack{i_1,\dots,i_k \\ j_1,\dots, j_k}} c_{i_1j_1}\dots c_{i_kj_k} \\
&\sum_{\substack{\text{directed trees with} \\ \text{k labeled edges}}} \sum_{\substack{\text{single trace} \\ \text{insertions}}}  \prod_{i=1}^{k+1} V_i \,
\label{res:free_e}
\end{split}
\ee
where $V_i$ is the planar connected correlator, given by (\ref{mm:cc}), on the $i$-th vertex on the tree, that contains the following operators: tr $a^{i_s}$ if the directed edge labelled $s$ leaves that vertex; tr $a^{j_s}$ if the directed edge labelled $s$ arrives at that vertex; any single trace operators inserted on that vertex. While in this work we will focus on the planar free energy, we have shown in previous work how to apply this approach to the planar limit of other observables, like the Wilson loop \cite{Fiol:2020bhf, Fiol:2020ojn} or extremal correlators \cite{Fiol:2021icm}.

The structure and main results of the paper are as follows. In section 2 we consider cases where the potential (\ref{def:MM_potential}) contains only a finite number of terms. For instance, for a potential with a finite number of single-trace terms, $V=\text{N} \left(c_4 \tr \phi^4+\dots +c_{2k} \tr \phi^{2k}\right)$ we obtain
\be
 {\cal F}_0(t)=\sum_{\substack{j_2,\dots, j_k  \\ j_2+\dots+j_k>0}} \frac{1}{j_2! \dots j_k!} \frac{(2j_2+\dots +k j_k-1)!}{(j_2+\dots +(k-1)j_k+2)!} (-x_2)^{j_2}\dots (-x_k)^{j_k} \, ,
\ee
where $x_i =\frac{(2i)!}{(i-1)!i!}c_{2i} t^i$. This result was recently derived in \cite{Anninos:2020geh} by different methods.

As a second example, for the potential $V=c_{2k2k} \tr \phi^{2k}\tr \phi^{2k}$ we obtain
\be
{\cal F}_0(y_k)=\sum_{m=1}^\infty (-y_k)^m \frac{(m-1)!}{(2m)!} B_{2m,m+1} (1z_1,2z_2,3z_3,\dots, m z_m) \, ,
\ee
where $z_j$ is defined as\footnote{Note that $z_n$ depends on $k$, but since we use it only for potentials with a single value of $k$, we don't make explicit this dependence in the notation.}
\be
z_n =  \frac{(nk-1)!}{(n(k-1)+2)!} \, ,
\label{defzn}
\ee
$y_k=2 \frac{(2k)!^2}{(k-1)!^2 k!^2} c_{2k 2k} t^{2k}$ and $B_{n,k}$ are partial Bell polynomials \cite{bell}. In this last case, we haven't been able to determine analytically the radius of convergence, but we have found an analytic bound, and have proved that at large $k$, the radius of convergence tends to $t_c\rightarrow 1/4$.

In section 3, we turn our attention to various examples of potentials (\ref{def:MM_potential}) with infinitely many terms, appearing in  four dimensional ${\cal N}=2$ supersymmetric gauge theories on $S^4$ upon localization. We first review the case of ${\cal N}=2$ SU(N) with 2N multiplets in the fundamental representation of the gauge group. Being a superconformal theory, the matrix model contains only double-trace terms in the planar limit. Then, we let go of conformality to consider two examples of non-conformal $\mathcal{N}=2$ theories. We show that their matrix models contain both single and double traces contributing to the planar limit. They thus fall in the category of theories for which (\ref{res:free_e}) applies. We compute their planar free energy and comment on the convergence of the resulting perturbative series. 

The appendix contains an alternative derivation of the results of section 2, using the more traditional approach. We derive the corresponding eigenvalue densities (or more precisely, their relevant moments). Besides taking more effort to derive, even after one obtains the eigenvalue densities, evaluating the planar free energies involves non-trivial mathematical identities, so it looks unlikely to us that one could have originally derived these planar free energies following this route, without knowing already the results.

This work can have a number of applications, and suggests a number of possible extensions. A promising application involves revisiting the derivation of exact glueball superpotentials  in four-dimensional ${\cal N}=1$ gauge theories, from the planar limit of an auxiliary matrix model \cite{Dijkgraaf:2002dh}. As for extensions, \cite{tutte, Gopakumar:2012ny} obtained the planar connected correlator for an arbitrary number of even operators, and up to two odd operators. For this reason, in this paper we have restricted ourselves to even potentials; the results in \cite{Bouttier:2002iw} might allow to extend our work to include also odd potentials. A more interesting open question, on which we are currently working on, is to extend our method to subleading 1/N terms in (\ref{freefdef}). Finally, the techniques used here for Hermitian one-matrix models have been applied so far to very specific multi-matrix models \cite{Fiol:2020ojn}; it would be interesting to extend them to generic multi-matrix models.

\section{Potentials with a finite number of terms}
In this section, we apply the approach described in the introduction to various examples of potentials with a finite number of terms. In all of these examples, we find the all-order perturbative series for the planar free energy in closed form.

\subsection{Potential with one single-trace term}
As a warm-up, consider first the case of a potential with just one single-trace term, $V=\text{N} c_{2k}\tr \phi^{2k}$, besides the Gaussian quadratic term. The planar free energy is then the generating function of planar connected correlators, and in this case all $k_i=k$ so using (\ref{samepowers}) and recalling the definition of $z_n$ (\ref{defzn}), we arrive at
\be
{\cal F}_0(t)= \sum_{n=1}^\infty \frac{(-c_{2k})^n}{n!} \langle  \left( \text{tr }\phi^{2k} \right)^n \rangle_c=
\sum_{n=1}^\infty  \frac{(nk-1)!}{n! (nk-n+2)!} \left (- x_k\right)^n=  \sum_{n=1}^\infty \frac{z_n}{n!} \left (- x_k\right)^n \, ,
\label{planarfreesingle}
\ee
where we have introduced the natural expansion parameter 
\be
x_k=\frac{(2k)!}{(k-1)!k!}c_{2k} t^k \, .
\label{singlex}
\ee
For the case $k=2$, this result appears at the end of \cite{Brezin:1977sv} while the general case appears in \cite{Anninos:2020geh}. The power series in (\ref{planarfreesingle}) are actually hypergeometric functions
\be
{\cal F}_0(t)=
-\frac{x_k}{k(k+1)}\, \, { }_{k+1}\text{F}_k \left[\begin{matrix} 1 & 1 & \frac{k+1}{k} &\dots & \frac{2k-1}{k} \\ &  2 & \frac{k+2}{k-1} &\dots & \frac{2k}{k-1} \end{matrix};  -\frac{k^k x_k}{(k-1)^{k-1}} \right] \, .
 \ee
It follows from the hypergeometric representation that the radius of convergence is $x_c=\frac{(k-1)^{k-1}}{k^k}$.  In view of (\ref{singlex}), the radius of convergence of the original 't Hooft coupling $t$ tends to $t_c \rightarrow 1/4$ when $k\rightarrow \infty$. 

\subsection{Potential with a finite number of single-trace terms}
Consider now the potential $V=\text{N} (c_4 \tr \phi^4 +\dots + c_{2k} \tr \phi^{2k})$. Matrix models with these potentials were studied in \cite{Bessis:1980ss} by different methods. In the context of 2d quantum gravity, these models gained further relevance after the work of \cite{Kazakov:1989bc}: in the double scaling limit \cite{Douglas:1989ve, Brezin:1990rb, Gross:1989vs} they reproduce the $(2,2k-1)$ minimal models coupled to quantum gravity. A different context where the planar limit of these matrix models plays a crucial role is in the computation of the exact glueball superpotential of ${\cal N}=1$ 4d gauge theories \cite{Dijkgraaf:2002dh}. Very recently these models have been revisited in \cite{Anninos:2020geh}, where they also deduce the planar free energy using different methods.

Again, since there are no double-trace terms, the planar free energy is the generating function of connected correlators. It is convenient to write the connected correlators in terms of the multiplicities $j_{k}$ of the operators $\tr \phi^{2k}$,
\be
\begin{split}
{\cal F}_0(t)=\sum_{m=1}^\infty \frac{(-1)^m}{m!} \sum_{p_1,\dots,p_m} c_{2p_1}\dots c_{2p_m}\langle \tr \phi^{2p_1} \dots \tr \phi^{2p_m}\rangle_c=\\
\sum_{m=1}^\infty \frac{(-1)^m}{m!} \sum_{\substack{j_2,\dots, j_k  \\ j_2+\dots+j_k=m}} \frac{m!}{j_2!\dots j_k!} c_4^{j_2}\dots c_{2k}^{j_k}  \langle \left(\tr \phi^{4}\right)^{j_2} \dots  (\tr \phi^{2k} )^{j_k} \rangle_c \, .
 \end{split}
 \ee
Then, recalling the definition of the couplings $x_k$ (\ref{singlex}) and using (\ref{mm:cc}) we arrive at a rather compact expression for the planar free energy of these models
\be
 {\cal F}_0(t)=\sum_{\substack{j_2,\dots, j_k  \\ j_2+\dots+j_k>0}} \frac{1}{j_2! \dots j_k!} \frac{(2j_2+\dots +k j_k-1)!}{(j_2+\dots +(k-1)j_k+2)!} (-x_2)^{j_2}\dots (-x_k)^{j_k} \, .
 \label{freemultisingle}
\ee
As a simple check, when only one of the terms in the potential is different from zero, (\ref{freemultisingle}) reduces to (\ref{planarfreesingle}). As a first non-trivial example, when $V=\text{N} (c_4 \tr \phi^4+c_6 \tr \phi^6)$, eq. (\ref{freemultisingle}) reduces to
\be
{\cal F}_0(t)= \sum_{m=1}^\infty  \sum_{j=0}^m \frac{ (3m-j-1)!}{j! (m-j)! (2m-j+2)! } (-x_2)^j (-x_3)^{m-j} \, ,
\ee
which reproduces the result in \cite{Anninos:2020geh}. It can be further rewritten as
\be
{\cal F}_0(t)=\sum_{m=1}^\infty (-x_3)^m \frac{(3m-1)!}{m! (2m+2)!} F\left( -2-2m,-m; 1-3m;-\frac{x_2}{x_3}\right) \, ,
\ee
which is simpler than the similar expression that appears in \cite{Anninos:2020geh}. As a second example, for $k=4$ eq. (\ref{freemultisingle}) reduces to
\be
{\cal F}_0(t) =\sum_{\substack{j_2,j_3,j_4 \\  j_2+j_3+j_4>0}} \frac{1}{j_2!j_3!j_4!} \frac{(2j_2+3j_3+4j_4-1)!}{(j_2+2j_3+3j_4+2)!} (-x_2)^{j_2} (-x_3)^{j_3} (-x_4)^{j_4} \, ,
\ee
which upon expansion reproduces eq. (4.27) in \cite{Anninos:2020geh}.


\subsection{Potential with one double-trace term}
We now switch to examples of potentials with double-trace terms. As far as we are aware, the first appearance of such potentials was in \cite{Das:1989fq}, in the context of 2d quantum gravity, where they were introduced to take into account higher order curvature effects, see also \cite{Korchemsky:1992tt, AlvarezGaume:1992np,Klebanov:1994pv, Klebanov:1994kv}. Matrix models with double trace terms have been considered in the computation of glueball superpotential of 4d ${\cal N}=1$ gauge theories in \cite{Balasubramanian:2002tm}.

The first example that we will consider is the potential with just one such term $V= c_{2k \, 2k} \tr \phi^{2k} \tr \phi^{2k}$. As explained in \cite{Fiol:2020bhf}, for potentials with just double traces, at order $m$ in the perturbative expansion of the planar free energy, we have to sum over all the ways to distribute $2m$ operators into $m+1$ connected correlators, such that the same connected correlators don't appear as products of lower order terms. It was further argued in \cite{Fiol:2020bhf} that this sum can be represented as a sum over tree graphs with $m$ edges. As we will show now, in the particular case at hand, since there is just one type of operator, $\tr \phi^{2k},$ the sum over trees simplifies drastically, and the contribution to the planar free energy is given at every order by a partial Bell polynomial.

To proceed, let's recall the basics of tree graphs (see \cite{harary,moon} for more details, or \cite{Fiol:2020bhf} for the bare minimum required in this work). A tree graph is a connected graph without loops. A tree with $m$ edges has $m+1$ vertices. To each vertex we associate its {\it degree} $d_i$: the number of vertices it is connected to. A simple result is that a set of $m+1$ numbers $(d_1,\dots,d_{m+1})$ is the degree sequence of a tree with $m+1$ vertices if and only if $\sum_{i} d_i =2m$.

In general, for a potential with just double traces, (\ref{res:free_e}) simplifies to \cite{Fiol:2020bhf}
\be
{\cal F}_0(t)= \sum_{m=1}^\infty \frac{(-1)^m}{m!} \sum_{\substack{i_1,\dots,i_m \\  j_1,\dots,j_m}} c_{i_1 j_1}\dots c_{i_m j_m}
\sum_{\substack{\text{directed trees with} \\ \text{m labeled edges}}} \prod_{i=1}^{m+1} V_i \, ,
\ee
where the product runs over the $m+1$ vertices of the tree. In the particular case when the potential contains just one double trace,
\be
{\cal F}_0(t)= \sum_{m=1}^\infty \frac{(-1)^m}{m!}  c_{2k 2k}^m 
\sum_{\substack{\text{directed trees with} \\ \text{m labeled edges}}}
\prod_{i=1}^{m+1} z_{d_i} \left( \frac{(2k)!}{(k-1)! k!}\right)^{d_i} t^{d_i k} \, .
\ee
Recalling that for a tree with $m+1$ vertices, $\sum_i d_i=2m$ \cite{moon}, and taking into account that for $m>1$ the directions of the arrows in the directed tree don't affect its contribution\footnote{As discussed in \cite{Fiol:2020bhf}, the case $m=1$ requires to be treated separately: in this case reversing the arrow in the edge does not change the tree, so we should not multiply by two. However, we are about to use the formula (\ref{countingtrees}) that counts the number of labeled trees for a given degree sequence. That formula, valid for $m>1$, is off by a factor $1/2$ when extended to $m=1$. Thus, these two factors cancel each other, and the final results we present are also valid for $m=1$.}
\be
{\cal F}_0(t)= \sum_{m=1}^\infty \frac{1}{m!} \left(-2 \frac{(2k)!^2}{(k-1)!^2 k!^2} c_{2k 2k} t^{2k}\right)^m 
\sum_{\substack{\text{trees with} \\ \text{m labeled edges}}} 
\prod_{i=1}^{m+1} z_{d_i} \, .
\ee
We define 
\be
y_k \equiv 2 \frac{(2k)!^2}{(k-1)!^2 k!^2} c_{2k 2k} t^{2k} \, ,
\label{doublek}
\ee
as the natural expansion parameter for these models. Next, given a tree, we denote by $\alpha_j$ the number of vertices with degree $j$. Then we make use of the fact that for $m>1$ the number of trees with $m$ labeled edges and a given degree sequence $(d_1,\dots,d_{m+1})$ is \cite{moon}\footnote{See the previous footnote for the case $m=1$.} 
\be
\frac{1}{m+1}  {\sum_j \alpha_j  \choose \alpha_1 \dots \alpha_{m}} {\sum_i (d_i-1) \choose d_1-1 \dots d_{m+1}-1} \, ,
\label{countingtrees}
\ee
to arrive at the following expression of the planar free energy as a sum over tree graphs
\be
{\cal F}_0(t)= \sum_{m=1}^\infty \frac{(-y_k)^m}{m(m+1)} \sum_{\substack{\text{degree sequences} \\ \text{for trees with m edges}}}
{\sum_i \alpha_i  \choose \alpha_1 \dots \alpha_{m}} \prod_{i=1}^{m+1} \frac{ z_{d_i}} {(d_i-1)! } \, .
\label{planarfreetrees}
\ee
The degree sequences in (\ref{planarfreetrees}) are partitions of $2m$ elements (the total amount of operators) into exactly $m+1$ parts (the number of connected correlators), so the multiplicities satisfy $\alpha_1+\dots+\alpha_m=m+1$ and $1\alpha_1+2\alpha_2+\dots +m \alpha_m=2m$. The planar free energy then can be rewritten as
\be
{\cal F}_0(t)=\sum_{m=1}^\infty \frac{(-y_k)^m}{m(m+1)}  \sum_{\substack{\alpha_1+\dots+\alpha_m=m+1 \\ 1\alpha_1+\dots +m \alpha_m=2m}} \frac{(m+1)!}{\alpha_1! \dots \alpha_m!} \left( \frac{1z_1}{1!}\right)^{\alpha_1} 
\left( \frac{2z_2}{2!}\right)^{\alpha_2}...\left( \frac{m z_m}{m!} \right)^{\alpha_m} \, ,
\ee
If we now recall the definition of the partial Bell polynomial \cite{bell}
\be
B_{n,k} (x_1,\dots, x_{n-k+1})=  \sum_{\substack{\alpha_1+\dots+\alpha_{n-k+1}=k \\ 1\alpha_1+\dots  (n-k+1)\alpha_{n-k+1}=n}} 
\frac{n!}{\alpha_1!\dots \alpha_{n-k+1}!} \left(\frac{x_1}{1!}\right)^{\alpha_1}  \left(\frac{x_2}{2!}\right)^{\alpha_2} ...  \left(\frac{x_{n-k+1}}{ (n-k+1)!}\right)^{\alpha_{n-k+1} }
\ee
we realize that the planar free energy can be written in terms of partial Bell polynomials,
\be
{\cal F}_0(y_k)=\sum_{m=1}^\infty (-y_k)^m \frac{(m-1)!}{(2m)!} B_{2m,m+1} (1z_1,2z_2,3z_3,\dots, m z_m) \, .
\label{freebell}
\ee
This is a pleasantly compact expression. In hindsight, the appearance of partial Bell polynomials is not surprising. At order $m$, the planar free energy receives contributions from the different ways to group a set of $2m$ operators into $m+1$ connected correlators, subject to the constraints mentioned above. The partial Bell polynomial $B_{n,k}$ enumerates all the ways to group a set of $n$ elements into $k$ groups, explaining the appearance of $B_{2m,m+1}$ at order $m$.

As a first check, in the particular case of $k=1$ the Bell polynomials in (\ref{freebell}) can be evaluated
\be
B_{2m,m+1}\left(\frac{1!}{2},\frac{2!}{2},\dots \frac{m!}{2}\right)= \frac{1}{2^{m+1}} \frac{(2m-1)! (2m)!}{(m+1)! m! (m-1)!} \, ,
\ee
which gives the planar free energy
\be
{\cal F}_0(y_1)=\frac{1}{2} \sum_{m=1}^\infty \left(-\frac{y_1}{2}\right)^m \frac{(2m-1)!}{m! (m+1)!} \, .
\label{doubletwo}
\ee
This reproduces the result obtained in \cite{Das:1989fq} for the $\tr \phi^2 \tr \phi^2$ potential (see also the appendix). For arbitrary $k$, we can evaluate the first terms of (\ref{planarfreetrees}),
\be
\begin{split}
{\cal F}_0(y_k) = -\frac{y_k}{2 k^2 (k+1)^2}+\frac{y^2_k}{4 k^3 (k+1)^2} -\frac{y_k^3}{12} \left[ \frac{6}{4k^4 (k+1)^2} +\frac{2}{k^3 (k+1)^3}\right]\\
+\frac{y_k^4}{20}\left[ \frac{10}{8k^5 (k+1)^2}+\frac{5}{k^4 (k+1)^3}+\frac{5(4k-1)}{6k^4 (k+1)^4}\right]+\dots
\end{split}
\label{planartreeexpand}
\ee
or directly the first terms of (\ref{freebell}),
\be
\begin{split}
{\cal F}_0(y_k)=-\frac{y_k}{2 k^2 (k+1)^2}+\frac{y^2_k}{4 k^3 (k+1)^2}-\frac{(7k+3)y_k^3}{24 k^4 (1+k)^3} \\
+\frac{(23k^2+16k+3)y_k^4}{48 k^5 (1+k)^4}-\frac{(455k^3+405k^2+133k+15)y_k^5}{480 k^6(1+k)^5}+\dots
\end{split}
\label{freebellfirst}
\ee


We now want to point out a relation between the planar free energies of two of the models discussed so far, the potential with one single-trace term, and the potential with one double-trace term. The fact that the arguments of the Bell polynomials in the planar free energy of the double-trace model, eq. (\ref{freebell}), are the coefficients of the expression for the planar free energy of the single-trace model, eq. (\ref{planarfreesingle}), implies that these two planar free energies ${\cal F}_0^{dt}$ and ${\cal F}_0^{st}$ are actually related. Indeed, it follows from  (\ref{planarfreesingle}), (\ref{freebell}) and the Fa\`a di Bruno's formula that
\be
{\cal F}_0^{dt} (y_k) = \left. \sum_{m=1}^\infty (-y_k)^m \frac{1}{(m+1)!} \frac{d ^{m-1}}{d z^{m-1} }\left(\frac{d {\cal F}_0^{st} (-z) }{dz}\right)^{m+1} \right \rvert_{z=0} \, .
\ee
Using the Lagrange inversion theorem, this relation can be rewritten as 
\be
\frac{d {\cal F}_0^{dt} (-y_k)}{dy_k} =\frac{1}{2} \left(\frac{d {\cal F}_0^{st}(-z)}{dz}\right)^2 \, ,
\label{singledouble}
\ee
where $z(y_k)$ is obtained from inverting the equation
\be
y_k=\frac{z}{\frac{d {\cal F}_0^{st}(-z)}{dz} } \, .
\ee
Let us conclude by mentioning that the free energies of the double scaling limits of these two models are also related \cite{Klebanov:1994kv}. 

\subsubsection{Radius of convergence}
We now want to discuss the radius of convergence of the planar free energy of this model, eq. (\ref{freebell}). For $k=1$, we can apply the quotient criterion to (\ref{doubletwo}), and find that the radius of convergence is $t_c^2 =\frac{1}{16 c_{22}}$. 

\begin{figure}
\centering
\includegraphics[width=.5\textwidth]{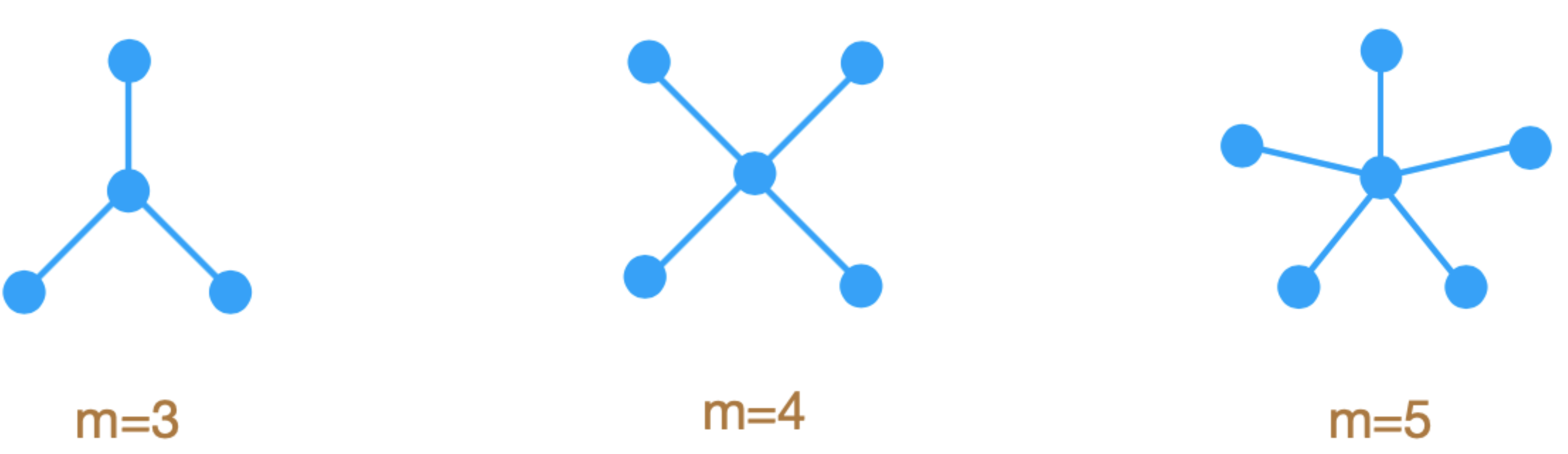}
\caption{Examples of star graphs with $m=3,4,5$ edges.}
\label{stargraphs}
\end{figure}

For any $k>1$, we can derive an upper bound on the radius of convergence by considering at every order just the contribution from a single tree, the star graph. The star graph with $m+1$ vertices is the single tree with $m$ vertices of degree 1 and one vertex of degree $m$, joined by edges to all the rest, see figure (\ref{stargraphs}). Its degree sequence is $\vec d =(1,\dots,1,m)$ and the multiplicities of the degrees are  $\vec \alpha=(m,0,\dots,0,1)$. In terms of planar connected correlators, this truncation amounts to, at order $m$, just consider the contribution from
\be
\langle (\tr \phi^{2k})^m \rangle_c \,\, \langle \tr \phi^{2k} \rangle^m \, ,
\label{starcont}
\ee
to the planar free energy. 

At a given order $m$, the contribution to the sum over trees by the star graph is
\be
{\cal F}_0^{\text{star}} (y_k)= \sum_{m=1}^\infty \frac{(mk-1)!}{m! (m(k-1)+2)!} \left(-\frac{y_k}{k(k+1)}\right)^m = \sum_{m=1}^\infty \frac{z_m}{m!} \left(-\frac{y_k}{k(k+1)}\right)^m \, .
\ee
This is exactly the same perturbative series as for the planar free energy of single trace models, eq. (\ref{planarfreesingle}), which is a consequence of the kind of contributions we are keeping, eq. (\ref{starcont}). If we truncate the sum over trees to just this contribution, it follows from the quotient criterion that the radius of convergence for this truncated series is
\be
c_{2k 2k} t_c^{2k}(\text{star})= \frac{ (k+1) (k-1)^{k-1} (k-1)!^2 k!^2}{2 k^{k-1} (2k)!^2} \, .
\ee
Notice that at large $k$, $t_c^{2k}(star)\rightarrow 1/4$. At a given order in $m$, all the other trees besides the star graph contribute, so the full coefficient is larger than the contribution coming from the star graph. The true radius of convergence is thus smaller or equal than the one obtained by truncating the sum to just the contribution from the star graph
\be
t_c^{2k}\leq t_c^{2k}(star) \, .
\ee

Finally, let's comment on the radius of convergene of (\ref{freebell}) as $k\rightarrow \infty$. The large $k$ limit of (\ref{freebell}) simplifies to

\be
{\cal F}_0(y_k)=\frac{1}{k^3}\sum_{m=1}^\infty \left(\frac{-y_k}{k}\right)^m \frac{(m-1)!}{(2m)!} B_{2m,m+1}(1^{1-2},2^{2-2},\dots,m^{m-2}) \, .
\label{largek}
\ee
We learn that in the large $k$ limit, the dependence on $k$ factors out, something that one can see  in the first terms of the expansion, eq. (\ref{freebellfirst}). It is possible to argue that the radius of convergence of (\ref{largek}) must be non-zero \cite{overflow}\footnote{We are very thankful to Max Alekseyev for providing this argument.}. Recall from (\ref{singledouble}) that the planar free energy for this model is obtained by series reversion from the planar free energy of the single-trace model. By Lagrange's inversion theorem, since the later series has a non-zero radius of convergence, so does the first. We can do better, and bound the radius of convergence of (\ref{largek}) noting that for $m$ large enough
\be
B_{2m,m+1}(1,2,\dots, m)<B_{2m,m+1}(1^{1-2},2^{2-2},\dots,m^{m-2}) < B_{2m,m+1}(1^{1-1},2^{2-1},\dots,m^{m-1}) \, .
\ee
The Bell polynomials at the ends can be evaluated  \cite{khelifa, abbas}, resulting in 
\be
{2m \choose m-1}m^{m-1}< B_{2m,m+1}(1^{1-2},2^{2-2},\dots,m^{m-2}) <  {2m-1 \choose m} (2m)^{m-1} \, ,
\ee
for $m$ large enough. From these bounds we conclude that the radius of convergence of (\ref{largek}), for large $k$ behaves as
\be
y_{k,c}= C k \hspace{1cm} \frac{1}{2e}\leq C \leq \frac{1}{e} \, .
\ee
The $C\leq 1/e$ bound coincides with the large $k$ limit of the bound obtained from the truncation to stargraph trees. It follows from (\ref{doublek}) that the radius of convergence for $t$ in the large $k$ limit tends to $\frac{1}{4}$. Interestingly, this is the same limit found for the model with just a single trace.


\subsection{Potential with one single and one double trace terms}

As a last example, we will consider a potential with both single- and double-trace terms. Specifically, take $V=\text{N} c_{2k}\, \text{tr } \phi^{2k}+ c_{2k 2k} \text{ tr }\phi^{2k} \text{tr }\phi^{2k}$. The advantage of considering the same power for the single-trace and the double-trace terms is that the sum over trees can again be simplified to yield Bell polynomials.

At order $m$, say we have $n$ double traces and $m-n$ single traces: there is a total of $2n+(m-n)=m+n$ operators, to be distributed into $n+1$ connected correlators, with the constraint that no operators coming from the same double trace sit in the same connected correlator, and no pair of correlators share operators from more than one double-trace. This translates into the combinatorial question of enumerating all trees with $n$ labelled edges, to which we add $m-n$ extra labelled vertices of degree 1 (called leaves in the graph theory literature). This enumeration is again given by a partial Bell polynomial, $B_{m+n,n+1}(1z1,\dots,m z_m)$, up to the overall normalization, so
\begin{equation}
{\cal F}(x_k,y_k)=\sum_{m=1}^\infty (-1)^m (m-1)! \sum_{n=0}^m \frac{x_k^{m-n} \, y_k^n }{(m-n)! (m+n)!} B_{m+n,n+1}(1z_1,\dots, mz_m) \, .
\label{mixedfree}
\end{equation}
When $c_{2k 2k}=0$, only the terms with $n=0$ survive, and (\ref{mixedfree}) reduces to (\ref{planarfreesingle}), while when $c_{2k}=0$, only the $n=m$ terms survive and (\ref{mixedfree}) reduces to (\ref{freebell}).

\section{Potentials with infinitely many terms}
Matrix models whose potential (\ref{def:MM_potential}) contains an infinite number of single an double trace deformations are 
relevant for two dimensional statistical physics models, Chern-Simons theories coupled to matter and models dual to M-theory backgrounds as discussed in \cite{Grassi:2014vwa}. More recently, it has been realized that supersymmetric localization \cite{Pestun:2007rz} reduces the evaluation of certain observables of Lagrangian ${\cal N}=2$ super Yang-Mills theories on $S^4$ to matrix models that can be recast in this form \cite{ Billo:2019fbi, Fiol:2020bhf}. Most of these works deal with Lagrangian ${\cal N}=2$ superconformal theories. In what follows we comment on possible implications of the analysis of the previous section for these family of models, and apply these techniques to derive the planar free energies on $S^4$ of a couple of non-conformal gauge theories, namely ${\cal N}=2^*$ SYM and $SU(N)$ with $\textnormal{N}_f<2\textnormal{N}$ multiplets in the fundamental representation.

Recall that supersymmetric localization allows to write the partition function of a ${\cal N}=2$ SYM theory on $S^4$ as the following matrix model \cite{Pestun:2007rz}

\begin{equation}
		Z_{S^4} (\tau_{\text{YM}})= \int da \, e^{-\frac{8\pi^2}{g_{\text{YM}}^2} \textnormal{Tr}(a^2)} \mathcal{Z}_{1-loop} (a) \, |\mathcal{Z}_{inst}(a,\tau)|^2 \, ,
\label{zs4}		
\end{equation} 
where $\mathcal{Z}_{inst}$ is the instanton contribution, that it is usually assumed to be negligible in the large N limit, and will be set to 1 one in what follows.  $\mathcal{Z}_{1-loop}$ is a factor arising from a $1-$loop computation in an auxiliary parameter, and itdepends on the choice of gauge group G and representations $R$ of the matter multiplet. It is given by products over the weights of the adjoint and matter representations $\alpha, w_R$ respectively
\be
\mathcal{Z}_{1-loop}= \prod_\alpha H(i\alpha \cdot \hat a) \prod_R \prod_{w_R} H(iw_R \cdot \hat a)^{-n_R} \, ,
\label{oneloopdet}
\ee
where H is related to the the Barnes G-function by $H(x)=G(1+x)G(1-x)$. For any ${\cal N}=2$ SYM theory with matter in representations with up to two indices, this 1-loop factor can be recast as an effective action involving single and double trace terms \cite{ Billo:2019fbi, Fiol:2020bhf}.
 
 Let's now comment on the convergence of the resulting perturbative series for the planar free energies. For conformal theories that admit a large N expansion, we expect that in the planar limit the perturbative series for observable quantities have a finite radius of convergence in the 't Hooft coupling \cite{Koplik:1977pf, tHooft:1982uvh}. In the case of ${\cal N}=2$ Lagrangian SCFTs, for observables that can be computed using supersymmetric localization, there is indeed strong numerical evidence that the radius of convergence of the perturbative planar series is given by $\lambda_c=\pi^2$ \cite{Beccaria:2020hgy, Beccaria:2021hvt}. As we will show, this value for the radius of convergence coincides with the large $k$ limit of the radius of convergence for both the model with a single-trace term and the model with a double-trace term discussed in the previous section.

On the other hand, for asymptotically-free non-conformal theories, we generically expect that the theory has renormalons \cite{tHooft:1977xjm, Parisi:1978az, Parisi:1978bj}. These renormalons can make every term in the 1/N expansion a divergent series \cite{DiPietro:2021yxb}. One could then envision exploring the appearance of renormalons for non-conformal ${\cal N}=2$ SYM using supersymmetric localization. However, it can be proven that for ${\cal N}=2$ SU(N) SYM on $S^4$,  observables that can be computed via supersymmetric localization are Borel summable at finite N \cite{Russo:2012kj, Aniceto:2014hoa, Honda:2016mvg}. Presumably, this implies that the perturbative planar series in $\lambda$ is also Borel summable. One can even speculate that they may be convergent. The explicit perturbative series that we derive later in this section pave the way for a numerical analysis of this possibility.

\subsection{Potential with infinitely many double-trace terms:  ${\cal N}=2$ SCFTs}
It was shown in \cite{ Billo:2019fbi, Fiol:2020bhf}  that for Lagrangian ${\cal N}=2$ superconformal field theories, the matrix model that one obtains from supersymmetric localization contains both single- and double-trace terms,
\be
\begin{split}
S_{int}=\sum_{n=2}^\infty \frac{\zeta_{2n-1} (-1)^n}{n}\left[(4-4^n)\alpha_G \tr a^{2n} +\beta_G \sum_{k=1}^{n-1} {2n \choose 2k} \tr a^{2n-2k} \tr a^{2k}\right. \\
\left. +\gamma_G \sum_{k=1}^{n-2} {2n \choose 2k+1} \tr a^{2n-2k-1} \tr a^{2k+1} \right] \, ,
\label{conformalaction}
\end{split}
\ee
where $\alpha_G,\beta_G,\gamma_G$ depend on the matter content of the theory. In theories where the matter in the fundamental representation scales with the rank of the gauge group, $\beta_G \neq 0$ in the planar limit. For those theories,
 only double-trace terms with even operators contribute to the planar limit, so for the purpose of studying the planar limit one can take \cite{Fiol:2020bhf}
\be
S_{int}=\beta_G \sum_{i,j=1}^\infty \frac{\zeta_{2i+2j-1}}{i+j} (-1)^{i+j} {2i+2j \choose 2i} \tr a^{2i} \tr a^{2j} \, .
\ee
In the previous section, we argued that for double-trace potentials with coefficients $c_{2k 2k}$ that don't scale with $k$, the radius of convergence tends to $1/4$. In the case at hand, if we focus on terms with the same trace, they are essentially of the form
\be
V = {4k \choose 2k} \tr \phi^{2k} \tr \phi^{2k} \, ,
\label{doublescale}
\ee
so the coefficient $c_{2k 2k}$ scales like $4^{2k}$ at large $k$. Recalling the relation between $y_k$ and $t$ for these potentials, eq. (\ref{doublek}), this simply implies that now $4t_c =1/4$ for large $k$, or equivalently $t_c=1/16$ in this limit. Comparing the kinetic terms in the original matrix model and in (\ref{zs4}), we learn that the 't Hooft coupling of the matrix model $t$ and the Yang-Mills 't Hooft coupling $\lambda=g_{\text{YM}}^2 \textnormal{N}$ are related by $t=\frac{\lambda}{16\pi^2}$. So for the potential (\ref{doublescale}) with just one such term, at large $k$ the radius of convergence tends to $\lambda_c =\pi^2$. This is actually the radius of convergence found for various planar series (or truncations thereof) of conformal ${\cal N}=2$ theories: it is the radius of convergence of the dispersion relation for the ${\cal N}=4$ magnon \cite{Beisert:2004hm}. It is also the radius of convergence analytically found in \cite{Fiol:2020bhf} for the (uncontrolled) truncation of the planar free energy of ${\cal N}=2$ SYM theories on $S^4$ with $\beta_G\neq 0$ to terms with just one value of $\zeta$. Finally, it is also the value found numerically in \cite{Beccaria:2020hgy, Beccaria:2021hvt} as the radius of convergence  for the planar free energy of (\ref{conformalaction}) when $\beta_G=0$, which corresponds to theories where the number of matter multiplets in the fundamental representation does not scale with N. We conjecture that for all observables of ${\cal N}=2$ SYM conformal  theories captured by supersymmetric localization, the radius of convergence of the planar limit is $\lambda_c = \pi^2$. There is also evidence for this being the radius of convergence of some perturbative series for the non-planar terms in the 1/N expansion \cite{Beccaria:2021ksw} and perhaps even non-conformal theories, as discussed below. The observation presented above, that this value coincides with the large $k$ limit of the radius of convergence of (\ref{doublescale}) certainly does not constitute a proof of this conjecture, but we expect this observation to play an important role in a (yet to be developed) full analytic proof.

\subsection{Potential with infinitely many single- and double-trace terms: ${\cal N}=2$ nonconformal theories}
We consider now two specific examples of non-conformal ${\cal N}=2$ super Yang-Mills theories. We will see that in both cases, the partition function obtained from supersymmetric localization can be rewritten as a matrix model with a potential with infinitely many single and double trace terms. Opposite to what happens in the conformal case, eq. (\ref{conformalaction}), now the single-trace terms have the right large N scaling to contribute to the planar limit. These two examples thus fall in the category of matrix models solved in the planar limit by (\ref{res:free_e}). 

\subsubsection{${\cal N}=2$ SU(N) with $\textnormal{N}_f<2\textnormal{N}$}
As a first example of non-conformal ${\cal N}=2$ theory, we will consider ${\cal N}=2$ SU(N) SYM with $\textnormal{N}_f<2\textnormal{N}$ multiplets in the fundamental representation. The corresponding 1-loop determinant (\ref{oneloopdet}) is
\be
Z_{1-loop}= \frac{\prod_{1=u<v}^N H(ia_u-ia_v)^2}{\prod_{u=1}^N H(ia_u)^{N_f}} \, .
\ee
Taking into account the perturbative expansion of the logarithm of the $H$ function
\be
\log H(x)= -(1+\gamma)x^2 -\sum_{n=2}^\infty \zeta_{2n-1} \frac{x^{2n}}{n} \, ,
\label{expandh}
\ee
the effective action can be rewritten as
\be
\begin{split}
S_{int} = &\textnormal{N}  \left[\left(\frac{N_f}{N}-2\right)(1+\gamma) \textnormal{tr}a^2 + \sum_{n=2}^\infty  \frac{\zeta_{2n-1}(-1)^n}{n} \textnormal{tr}a^{2i} \right] \\
&+\sum_{n=2}^\infty \frac{\zeta_{2n-1} (-1)^n}{n}\left[\sum_{k=1}^{n-1} {2n \choose 2k} \tr a^{2n-2k} \tr a^{2k} + \sum_{k=1}^{n-2} {2n \choose 2k+1} \tr a^{2n-2k-1} \tr a^{2k+1} \right] \, .
\end{split}
\ee 
While this potential has single-trace terms, they are all even. Following the arguments in \cite{Fiol:2021icm}, it follows that the double-trace terms with odd powers don't contribute to the planar free energy. Thus, for the purpose of computing the planar free energy we can restrict to
\be 
\begin{split}
S_{int} = &\textnormal{N}  \left[\left(\frac{\textnormal{N}_f}{\textnormal{N}}-2\right)(1+\gamma) \textnormal{tr}a^2 + \sum_{n=2}^\infty  \frac{\zeta_{2n-1}(-1)^n}{n} \textnormal{tr}a^{2i} \right] + \sum_{i,j} \frac{\zeta_{2i+2j-1}(-1)^{i+j}}{i+j} {2i + 2j \choose 2i} \textnormal{tr}a^{2i} \textnormal{tr}a^{2j} \, ,
\end{split}
\ee
so (\ref{zs4}) can be rewritten as a Hermitian matrix model with a potential with an infinite number of single and double trace terms. The single-trace terms appearing in the potential are all proportional to the beta function of the theory, so they vanish in the particular case $\textnormal{N}_f=2\textnormal{N}$, the conformal case. On the other hand, the double-trace terms are those of the conformal case. Note also that as long as $2\textnormal{N}-\textnormal{N}_f$ scales like N in the large N limit, the single-trace terms contribute to the planar limit. As an illustration, we apply (\ref{res:free_e})  truncating the expansion of the planar free energy to the terms that contain only one zeta function
\be 
\begin{split}
\mathcal{F} = &-\left( \frac{\textnormal{N}_f}{\textnormal{N}}-2 \right)(1+\gamma) \left(\frac{\lambda}{16\pi^2} \right) + \sum_{p=2}^\infty \frac{\zeta_{2p-1}}{p}\frac{(2p)!}{(p+1)!p!}\left(- \frac{\lambda}{16\pi^2} \right)^p \\
& - \sum_{i,j=1}^\infty \frac{\zeta_{2i+2j-1}}{(i+j)} \left(\frac{-\lambda}{16\pi^2} \right)^{i+j} {2i+2j \choose 2i} \frac{(2i)!(2j)!}{(i+1)!i!(j+1)!j!} + \cdots \, ,
\end{split}
\ee

\subsubsection{${\cal N}=2^*$}
As our second example of non-conformal gauge theory, let's consider ${\cal N}=2^*$ SU(N). This theory is the result of adding a mass term to the hypermultiplet of ${\cal N}=4$ SU(N) SYM.   ${\cal N}=2^*$ SU(N) has already been studied using supersymmetric localization \cite{Russo:2013kea}. The 1-loop determinant (\ref{oneloopdet}) is now
\be
Z_{1-loop}= \frac{\prod_{u<v} H(ia_u-ia_v)^2}{\prod_{u<v} H(ia_u-ia_v-M)H(ia_u-ia_v+M)} \, .
\ee
Taking the logarithm of the previous expression and recalling (\ref{expandh}), this can be rewritten - up to a constant term - as
\be
\begin{split}
& S_{int}= 2\textnormal{N} \sum_{j=1}^\infty \sum_{n=j+1}^\infty \frac{\zeta_{2n-1}}{n} {2n \choose 2j} (-1)^j M^{2n-2j} \text{tr }a^{2j} \\
&+\sum_{i,j=1}^\infty \left[ {2i+2j \choose 2i} (-1)^{i+j} \sum_{n=i+j+1} \frac{\zeta_{2n-1}}{n} {2n \choose 2i+2j} M^{2n-2i-2j}\right] \text{tr }a^{2i} \text{tr }a^{2j} \\
&+\sum_{i,j=1}^\infty \left[ {2i+2j+2 \choose 2i+1} (-1)^{i+j} \sum_{n=i+j+2} \frac{\zeta_{2n-1}}{n} {2n \choose 2i+2j+2} M^{2n-2i-2j-2}\right] \text{tr }a^{2i+1} \text{tr }a^{2j+1} \, ,
\end{split}
\ee
so again the matrix model coming from localization can be recasted as a matrix model with a potential with single and double trace terms, and the single-trace terms come with a power of N, so they contribute to the planar limit. It is thus possible to write the planar free energy for this theory on $S^4$ using (\ref{res:free_e}). In particular, the terms with a single value of $\zeta$ are
\be
{\cal F}_0= -\sum_{p=1}^\infty \frac{2 (2p)! (2p+1)!}{p! p! (p+1)! (p+2)!} \left(\frac{-\lambda}{16 \pi^2}\right)^p \sum_{m=1}^\infty \frac{\zeta_{2m+2p-1}}{m+p} {2m+2p \choose 2p} M^{2m}+\dots
\ee
where the dots stand for terms with two or more values of $\zeta$. It is possible to rewrite this result in integral form, which allows to explore the large $\lambda$ regime. Upon performing the sums we obtain
\be 
\mathcal{F}_0 = - \frac{4\pi^2}{\lambda} \int_0^\infty \textnormal{d}w \frac{\sinh^2 (w M)}{w^3 \sinh^2 w} \left( J_1\left( \frac{w \sqrt{\lambda}}{\pi} \right)^2 - \frac{w^2 \lambda}{4\pi^2}\right) \, .
\ee 
with $J_1$ a Bessel function. As a check, if we keep only the $M^2$ term in the expression above, we reproduce the result of \cite{Russo:2013kea}. For this truncation at order $M^2$, it was proven in  \cite{Russo:2013kea} that the radius of convergence is again $\lambda_c=\pi^2$. Because the theory is no longer conformal, the coupling runs, and this coupling should be understood as evaluated at the scale given by the radius of $S^4$. The result of \cite{Russo:2013kea} implies that at first order in conformal perturbation theory, the radius of convergence remains the same as the one found in the conformal cases reviewed above. It will be interesting to determine whether this is still the case for the full planar series.

\acknowledgments
We would like to thank Max Alekseyev for answering our questions about Bell polynomials \cite{overflow} and Beatrix M\"uhlmann for thoroughly reading and commenting a previous draft of this paper. We acknowledge helpful conversations with Dionysios Anninos, Marcos Mari\~no and Jorge Russo.
Research supported by  the State Agency for Research of the Spanish Ministry of Science and Innovation through the ``Unit of Excellence Mar\'ia de Maeztu 2020-2023'' award to the Institute of Cosmos Sciences (CEX2019-000918-M) and PID2019-105614GB-C22, and by AGAUR, grant 2017-SGR 754.  A. R. F. is further supported by an FPI-MINECO fellowship. 

\appendix
\section{Saddle point analysis}

In this appendix we reproduce to some extent the results we found in the first section, for potentials with finitely many terms, using the methods introduced in \cite{Bessis:1980ss}. \cite{Bessis:1980ss} considered only potentials with single-trace terms, and the extension to potentials with double-trace terms was worked out in \cite{Cicuta:1990uc, Grassi:2014vwa}. We follow  \cite{Grassi:2014vwa} closely. The matrix model considered is
\be
V(M)=\frac{1}{2g} \tr M^2 + \textnormal{N} \sum_k c_{2k} \tr M^{2k} +\sum_{jk} c_{2j\, 2k} \tr M^{2j} \tr M^{2k} \, .
\ee
After diagonalization of the matrix $M$, introduce the density of eigenvalues $\rho(\lambda)$, and its moments
\be
\rho_k = \int d\lambda \rho(\lambda) \lambda^k \, .
\ee
One of the basic quantities is $R_0(\xi,t)$, defined as the positive solution of
\be
\xi= \frac{1}{t}R_0 +\sum_{k\geq 2}  b_kR_0^k \, ,
\label{bigrdef}
\ee
where $\xi$ can be thought of as an auxiliary variable and
\be
b_{k} = \frac{(2k)!}{k!(k-1)!}\left(c_{2k}+2\sum_j c_{2j,2k} \rho_{2j}\right) \, .
\label{tildea}
\ee
Then, the planar free energy (after subtracting the Gaussian term) is given by
\be
{\cal F}_0(t)= \int_0^1 d\xi (1-\xi) \log \frac{R_0(\xi,t)}{t\xi} +\sum_{j,k} c_{2j,2k} \rho_{2j}\rho_{2k} \, .
\label{freeapp}
\ee

The starting point of our approach is to solve (\ref{bigrdef}) by means of the Lagrange inversion theorem. In the case of potentials with just single-trace terms, this already yields an explicit expression for $R_0(\xi,t)$ and we can proceed to evaluate the planar free energy (\ref{freeapp}). The case of potentials with double-trace terms is {\it a priori} more complicated, since the coefficients $b_k$ now depend on the moments of the eigenvalue density - see (\ref{tildea}) - which at this stage is not known explicitly. In this case, one can further relate the eigenvalue density moments to $R_0$ through

\be
\rho_{2l} =\frac{(2l)!}{l!^2} \int_0^1 d\xi R_0^{l} \, ,
\label{theroint}
\ee
and this is enough to determine $\rho_{2k}$ and $R_0$.

To proceed, define
\be
g(x)=\frac{1}{t}+\sum_{k\geq 2} b_k x^{k-1} \, ,
\ee
so according to Lagrange's inversion formula
\be
R_0(\xi,t)= \sum_{n=0}^\infty \frac{\xi^{n+1}}{(n+1)!} \frac{d^n}{d x^n} \frac{1}{g(x)^{n+1}} \Big\rvert_{x=0} \, .
\ee
The first terms of the perturbative expansion of $R_0(t,\xi)$ are
\be
\frac{R_0}{\xi t}= 1-b_2 t^2 \xi +(2b_2^2t^4-b_3t^3)\xi^2+(-b_4t^4+5b_2b_3t^5-5b_2^3t^6)\xi^3+\dots
\label{pertr}
\ee
and the coefficients that appear in this expansion constitute the integer sequence A111785 in \cite{oeisa111785}.
Let's consider as a first application the case of the potential with finitely many single-trace terms. In this case, the functions $b_k$ reduce to $b_kt^k=x_k$, with $x_k$ defined in (\ref{singlex}), so the first terms of the perturbative expansion of $R_0(t,\xi)$ are
\be
\frac{R_0}{\xi t}= 1-x_2 \xi +(2x_2^2-x_3)\xi^2+(-x_4+5x_2x_3-5x_2^3)\xi^3+
(14x_2^4-21x_2^2x_3+3x_3^2+6x_2x_4-x_5)\xi^4
\dots
\label{pertrx}
\ee
Carrying out the integral for the planar free energy (\ref{freeapp}) we find 
\be
{\cal F}_0(x_i)= -\frac{x_2}{6}+\frac{x_2^2}{8}-\frac{x_2^3}{6}+\frac{7 x_2^4}{24}-\frac{x_3}{12}+\frac{x_2x_3}{5}-\frac{x_2^2x_3}{2}+\frac{x_3^2}{12}-\frac{x_2 x_3^2}{2}-\frac{x_4}{20}+\dots
\ee
in agreement with the first terms in the expansion of the expression we found in the main text, eq. (\ref{freemultisingle}). The check we have just performed has an important consequence: it gives an expression for the integral in (\ref{freeapp}) in the general case. To understand why, notice that the perturbative series for $R_0(\xi,t)$ in the general case, (\ref{pertr}), and in the particular case of just single-trace terms, (\ref{pertrx}), are the same, just with the substitution $b_k t^k\rightarrow x_k$. Therefore, the outcome of the integral in (\ref{freeapp}) for the general case is the same as for the particular case, with the substitution $b_k t^k\rightarrow x_k$. In the particular case, the planar free energy (\ref{freeapp}) is just given by the the first term since $c_{2j,2k}=0$, so it must coincide with the result found in the main text (\ref{freemultisingle}). In summary, we learn that
\be
\int_0^1 d\xi (1-\xi) \log \frac{R_0(\xi,t)}{t\xi} =
\sum_{\substack{j_2,\dots, j_k  \\ j_2+\dots+j_k>0}} \frac{1}{j_2! \dots j_k!} \frac{(2j_2+\dots +k j_k-1)!}{(j_2+\dots +(k-1)j_k+2)!} (-b_2 t^2)^{j_2}\dots (-b_k t^k)^{j_k} 
\label{therint}
\ee


Let's now move to the case of potentials with just one double-trace term. The $\tr \phi^2 \tr \phi^2$ case can solved completely
\be
R_0(\xi,t)= 2\xi t \frac{\sqrt{1+16c_{22}t^2}-1}{16c_{22}t^2} \, ,
\ee
\be
\rho_{2} (t)=  \frac{\sqrt{1+16c_{22}t^2}-1}{8c_{22}t} \, ,
\ee
\be
{\cal F}_0(t)= \frac{1}{2} \log \left( \frac{\sqrt{1+16c_{22}t^2}-1}{8c_{22}t^2}\right)+c_{22} \rho_2^2 \, .
\ee
This reproduces eqs. (32)-(34) of \cite{Das:1989fq}. For generic $\text{tr }\phi^{2k} \text{tr }\phi^{2k}$, the equation (\ref{bigrdef}) for $R_0$ simplifies to
\be
\xi= \frac{1}{t}R_0+b_k R_0^k \, ,
\label{bigrdouble}
\ee
which leads to
\be
R_0(\xi,t)= \xi t \sum_{m=0}^\infty \frac{ (km)! }{m! (m(k-1)+1)!  }\left(- \xi^{k-1} b_k t^k \right)^m \, .
\ee
We can carry out the integral in (\ref{theroint}), and we arrive at an implicit equation for the density moment $\rho_{2k}(t)$
\be
\rho_{2k}(t)= -\frac{(2k)!}{k! (k-1)!} t^k \sum_{m=1}^\infty \frac{z_m}{ (m-1)!} (-b_k t^k)^{m-1} \, .
\label{rodouble}
\ee
If we define
\be
\bar \rho = \frac{k! (k-1)!}{(2k)! t^k}\rho_{2k} \, ,
\ee
this equation can be rewritten
\be
\bar \rho=\sum_{m=1}^\infty \frac{z_m}{ (m-1)!} (-y_k \bar \rho)^{m-1}
\label{barroimp}
\ee
Applying the Lagrange inversion formula to (\ref{barroimp}), we obtain an explicit expression for $\rho_{2k}(y_k)$
in terms of partial Bell polynomials
\be
\rho_{2k}(y_k)=\frac{(2k)!}{k! (k-1)!} t^k \sum_{n=0}^\infty (-y_k)^n \frac{n!}{(2n+1)!} B_{2n+1,n+1}(1z_1,\dots, (n+1)z_{n+1}) \, .
\ee
As a check, for $k=1$ this reduces to the result for $\rho_2$ quoted above. From (\ref{therint}), we deduce that in this case
\be
{\cal F}_0 (y_k)=\sum_{m=1}^\infty \frac{z_m}{m!}(-y_k\bar \rho)^m+\frac{1}{2}y_k \bar \rho^2 \, .
\ee
Taking the derivative against $y_k$ we deduce that
\be
-\frac{d {\cal F}_0(y_k)}{dy_k}= \frac{1}{2}  \bar \rho^2 \, ,
\ee
so the equivalence of both methods amounts to the mathematical identity
\be
\sum _{m=0}^\infty u^m \frac{(m+1)!}{(2m+2)!} B_{2m+2,m+2}(x_1,\dots,x_{m+1})=
\frac{1}{2} \left( \sum_{n=0}^\infty u^n \frac{n!}{(2n+1)!} B_{2n+1,n+1}(x_1,\dots, x_{n+1}) \right)^2
\ee
that can be proven by manipulating the generating function of Bell polynomials \cite{overflow}\footnote{We are very thankful to Max Alekseyev for providing this argument.} .

The last example that we will consider in this appendix is $V=\textnormal{N} c_{2k}\, \text{tr } \phi^{2k}+ c_{2k,2k} \text{ tr }\phi^{2k} \text{tr }\phi^{2k}$. The equation for $\rho_{2k}$ is still (\ref{rodouble}), which in terms of $\bar \rho$ is now
\be
\bar \rho (x_k,y_k)=\sum_{m=1}^\infty \frac{z_m}{(m-1)!} \left(-x_k-y_k\bar \rho  \right)^{m-1} \, ,
\label{mixedro}
\ee
whose solution is
\be
\bar \rho =\sum_{m=1}^\infty (-1)^{m+1} \sum_{n=1}^m \frac{x_k^{m-n}y_k^{n-1}}{(m-n)! (n+m-1)!} B_{n+m-1,n}(1z_1,\dots, m z_m) \, .
\label{mixedrobell}
\ee
Using (\ref{therint}), the planar free energy can be written as 
\be
{\cal F}_0(x_k,y_k)=\sum_{m=1}^\infty \frac{z_m}{m!}(-x_k-y_k \bar \rho)^m +\frac{1}{2} y_k \bar \rho^2 \, .
\ee
Taking partial derivatives with respect to $x_k$ and $y_k$ we find
\be
\frac{\partial {\cal F}_0}{\partial x_k}= -\bar \rho \, ,\hspace{1cm} \frac{\partial {\cal F}_0}{\partial y_k}=-\frac{1}{2}\bar \rho^2 \, .
\ee
Taking the derivative of the result we found in the main text, eq. (\ref{mixedfree}) against $x_k$, we do indeed recover eq. (\ref{mixedrobell}), proving the equivalence of both methods.

\end{document}